\documentclass[conference]{IEEEtran}
\IEEEoverridecommandlockouts
\usepackage{cite}
\usepackage{amsmath,amssymb,amsfonts}
\usepackage{epsfig}
\usepackage{theorem}
\usepackage{graphicx}
\usepackage{textcomp}
\usepackage{xcolor}
\def\BibTeX{{\rm B\kern-.05em{\sc i\kern-.025em b}\kern-.08em
    T\kern-.1667em\lower.7ex\hbox{E}\kern-.125emX}}

\DeclareMathOperator{\rect}{rect}

\begin{document}

\title{Continuous Phase Modulation of Phase Coded Transmit Waveforms using Multi-Tone Sinusoidal Frequency Modulation
\thanks{This research was funded by the Office of Naval Research grant \# N0001421WX00843.}
}

\author{\IEEEauthorblockN{David A. Hague}
\IEEEauthorblockA{\textit{Sensors and Sonar Systems Department} \\
\textit{Naval Undersea Warfare Center}\\
Newport, RI USA\\
david.a.hague@navy.mil}
}
\maketitle

\begin{abstract}
Phase Coded (PC) waveforms possess desirable Auto-Correlation Function (ACF) properties for use in radar and sonar systems.  However, their spectra possess high spectral leakage due to the abrupt phase transitions between the chips in the waveform.  This paper describes a method of Continuous Phase Modulation (CPM) to reduce a PC waveform's spectral leakage using the Multi-Tone Sinusoidal Frequency Modulation (MTSFM) model. The MTSFM-CPM model represents the PC waveform's instantaneous phase as a finite Fourier series.  This representation smooths the abrupt phase transitions between chips resulting in a spectrally compact waveform.  This smoothing of the PC waveform's instantaneous phase introduces perturbations to the waveform's ACF mainlobe and sidelobe structure.  Adjusting the MTSFM-CPM waveform's parameters refines its ACF mainlobe and sidelobe structure while also preserving its compact spectral shape.
\end{abstract}


\section{Introduction}
The vast majority of waveform diversity research over the last several decades has focused on the design of Phase Coded (PC) waveforms \cite{BluntIV}.  These waveforms are composed of a set of continguous sub-pulses, also known as chips, each possessing its own phase shift.  The set of phase-shifts, known as the phase code, is utilized as a discrete set of parameters that are selected to generate a waveform with distinct properties such as low sidelobes in its Ambiguity Function (AF) or Auto-Correlation Function (ACF) \cite{Levanon}.  Theoretically there exists a nearly endless combination of employable phase codes making PC waveforms an extremely versatile parameterized waveform model.  There continues to be extensive research on designing optimal PC waveforms for Multiple-Input Multiple-Output (MIMO) radar applications \cite{jianLiII, PalomarIII, RangaswamyI} and cognitive radar applications \cite{AubryII, AubryIII, PrabhuBabuII}.  Additionally, the developement of computationally efficient algorithms to design PC waveforms with specific ACF/AF properties is very much of interest to the radar and sonar communities \cite{AubryI, PrabhuBabuIII, PalomarI, PalomarII, SoltanalianI}.

In addition to possessing a desired AF/ACF shape and being constant modulus, waveforms are also generally required to be spectrally compact where the vast majority of the waveform's energy is densely concentrated in the system's operational band of frequencies.  Radar transmissions are often constrained by spectral masks which strictly limit the amount of waveform energy that resides outside the radar's allocated spectral band \cite{specMask}.  Additionally, the frequency response of the transmitter imparts spectral shaping of the physically emitted waveform.   The resulting transmitted signal therefore contains Amplitude Modulation (AM) and Phase Modulation (PM) effects which can perturb the waveform's spectral and AF/ACF shape characteristics \cite{Hague_GSFM_ASA, BluntI, BluntII}.  These deleterious effects can be minimized by transmitting waveforms that densely concentrate their energy in a band of frequencies where the transmitter's frequency response does not vary substantially.  


Most Frequency Modulated (FM) waveforms are quite spectrally compact.  This is largely because FM waveforms have modulation functions that are smooth, differentiable, and free of transient artifacts.  The spectra of PC waveforms however possess excessive spectral leakage outside their band of operation due to the abrupt phase transitions between chips \cite{Levanon}.   The PM and AM effects imposed by practical transmitters and their driving electronics distort the waveform transmitted into the medium thus corrupting its AF/ACF properties.  There are a number of Continuous Phase Modulation (CPM) methods which smooth the transition between chips by introducing continuity in the first several derivatives of the PC waveform's instantaneous phase \cite{BluntIII}.  This smoothing results in a more spectrally compact waveform while still retaining the waveform diversity afforded by phase coding \cite{BluntI}.  However, the degree of smoothing necessary to synthesize waveforms with high spectral compactness also introduces perturbations to the waveform's instantaneous phase.  These perturbations degrade the waveform's properties such as AF/ACF shape \cite{PCFM} necessitating re-optimization of the waveform's phase code \cite{BluntII}. 

This paper investigates another method of CPM for PC waveforms using Multi-Tone Sinusoidal Frequency Modulation (MTSFM) henceforth referred to as the MTSFM-CPM waveform.  The PC waveform's instantaneous phase is represented as a finite Fourier series via the MTSFM model.   This results in a waveform with an infinitely differentiable and therefore smooth instantaneous phase function which substantially improves the waveform's spectral compactness properties.  Reducing the number of harmonics in the representation increases the degree of smoothing of the PC waveform's instantaneous phase and therefore increases its spectral compactness.  While these more spectrally compact waveforms also possess degraded AF/ACF properties due to the perturbations inherent in the CPM process, the parameters of the MTSFM-CPM model can be modified to refine those AF/ACF properties.  This smoothing and re-optimization process results in waveforms with desireably low AF/ACF sidelobes and the spectral compactness properties necessary for transmission on practical transmitter devices and their driving electronics. 

\section{Waveform Signal Model}
\label{sec:sigMod}  

This paper assumes a monostatic radar/sonar system that transmits narrowband (i.e, small fractional bandwidth) waveforms.  A basebanded PC waveform with $N$ chips and pulse length $T$ defined over the interval $-T/2 \leq t \leq T/2$ is expressed as
\begin{equation}
s_{pc}\left(t\right) = \frac{1}{\sqrt{T}}\sum_{i=1}^N\rect\left[\dfrac{t-\left(i-1\right)t_b+\left(T-t_b\right)/2}{t_b}\right] e^{j\phi_i}
\label{eq:PC_Waveform}
\end{equation}
where $t_b = T/N$ is the chip duration, $\phi_i$ is the phase code of the PC waveform, and the amplitude term $1/\sqrt{T}$ normalizes the waveform to unit energy.  The general model for a unit energy Phase Modulated (PM) transmit waveform $s\left(t\right)$ is modeled as a complex analytic signal expressed as
\begin{equation}
s\left(t\right) = \dfrac{\rect\left(t/T\right)}{\sqrt{T}}e^{j\varphi\left(t\right)}
\end{equation}
where $\varphi\left(t\right)$ is the instantaneous phase of the waveform.  The waveform's corresponding modulation function maps its instantaneous frequency as a function of time
\begin{equation}
m\left(t\right) = \frac{1}{2\pi}\dfrac{\partial \varphi\left(t\right)}{\partial t}.
\end{equation}
A PC waveform's instantaneous phase can be interpreted as a PM waveform with a piecewise continuous instantaneous phase $\varphi_{pc}\left(t\right)$ expressed as
\begin{equation}
\varphi_{pc}\left(t\right) = \sum_{i=1}^N\rect\left[\dfrac{t-\left(i-1\right)t_b+\left(T-t_b\right)/2}{t_b}\right] \phi_i.
\label{eq:PC_Phase}
\end{equation}

One of the primary waveform design considerations for this paper is how densely the waveform concentrates its energy in a specified operational band of frequencies.  A commonly utilized method of measuring this property, referred to here as Spectral Compactness (SC), is defined as the ratio of waveform energy in a contiguous band of frequencies $\Delta f$ to the total energy of the waveform across all frequencies expressed as \cite{BluntIV, Hague_GSFM_ASA, Hague_AES} 
\begin{IEEEeqnarray}{rCl}
\Phi\left(\Delta f\right) = \dfrac{\int_{-\Delta f/2}^{\Delta f/2}|S\left(f\right)|^2df}{\int_{-\infty}^{\infty}|S\left(f\right)|^2df} = \int_{-\Delta f/2}^{\Delta f/2}|S\left(f\right)|^2df.
\label{eq:THETA}
\end{IEEEeqnarray}
where $S\left(f\right)$ is the waveform's Fourier transform.  Note that the second integral results from the assumption that the basebanded waveform's energy in the denominator is unity.  This SC measure is often utilized as it provides a fair means of comparison between FM waveforms and coded waveform techniques such as PC and Frequency Shift Keying (FSK) \cite{Hague_GSFM_ASA}. 

This paper assumes a Matched-Filter (MF) receiver is used to process target echoes.  The narrowband AF measures the correlation between a waveform and its Doppler shifted versions expressed as 
\begin{equation}
\chi\left(\tau, \nu\right) = \int_{-\infty}^{\infty} s\left(t-\frac{\tau}{2}\right) s^*\left(t+\frac{\tau}{2}\right) e^{j 2 \pi \nu t}dt.
\label{eq:AF}
\end{equation}
where $\nu$ is the Doppler frequency shift \cite{Rihaczek, Levanon}.  Of primary interest in this paper is the mainlobe and sidelobe structure of the zero-Doppler cut of the AF, the ACF, expressed as
\begin{equation}
R\left(\tau\right) = \int_{-\infty}^{\infty} s\left(t-\frac{\tau}{2}\right) s^*\left(t+\frac{\tau}{2}\right)dt.
\label{eq:ACF}
\end{equation}
A metric that accurately describes the ACF mainlobe structure is the area $A_0$ under the mainlobe region of the ACF which can be approximated as \cite{Rihaczek} 
\begin{IEEEeqnarray}{rCl}
A_0 = \int_{-\Delta \tau}^{\Delta \tau}|R\left(\tau\right)|^2 d\tau \cong \dfrac{\pi}{2\beta_{rms}}
\label{eq:mainlobe}
\end{IEEEeqnarray}
where $\Delta \tau$ denotes the first null of the ACF and therefore the null-to-null mainlobe width is $2\Delta \tau$.  The $\beta_{rms}$ term is the waveform's RMS bandwidth which for unit energy waveforms is expressed as 
\begin{IEEEeqnarray}{rCl}
\beta_{rms} = 2\pi\sqrt{\int_{-\infty}^{\infty}\left(f-f_0\right)^2|S\left(f\right)|^2 df}
\label{eq:rmsBand}
\end{IEEEeqnarray} 
where $f_0$ is the spectral centroid of $S\left(f\right)$.  

There are several metrics that describe the sidelobe structure of a waveform's ACF.  Two of the most common metrics are the Peak Sidelobe Level (PSL) and the Integrated Sidelobe Ratio (ISR).  The PSL is expressed as
\begin{IEEEeqnarray}{rCl}
\text{PSL} =  \dfrac{\text{max}\bigl\{|R\left(\tau\right)|\bigr\}_{\Delta \tau}^{T}}{\text{max}\bigl\{|R\left(\tau\right)|\bigr\}_{0}^{\Delta \tau}} = \text{max}\bigl\{|R\left(\tau\right)|\bigr\}_{\Delta \tau}^{T}.
\label{eq:PSL}
\end{IEEEeqnarray} 
Note that the rightmost expression in \eqref{eq:PSL} results from the assumption that the waveform is unit energy and thus the maximum value of $|R\left(\tau\right)|$ is unity which occurs at $\tau = 0$.  Another extensively utilized metric is the Integrated Sidelobe Ratio (ISR) which is the ratio of the area under the sidelobe region of $|R\left(\tau\right)|^2$ to the area under mainlobe region as shown in \eqref{eq:mainlobe} expressed as
\begin{IEEEeqnarray}{rCl}
\text{ISR}~=\dfrac{A_{\tau}}{A_0} = \dfrac{\int_{\Delta \tau}^{T}|R\left(\tau\right)|^2 d\tau}{\int_{0}^{\Delta \tau}|R\left(\tau\right)|^2 d\tau}.
\label{eq:ISR}
\end{IEEEeqnarray}
Note that the integration is performed only over positive time-delays since the ACF is even-symmetric in $\tau$.

Another sidelobe metric which is gaining increasing use for waveform optimization is the Generalized Integrated Sidelobe Ratio (GISR) \cite{bluntGISR}.  The GISR generalizes the ISR metric in \eqref{eq:ISR} by evaluating a $\ell_p$-norm \cite{prabhuBabuLp1, prabhuBabuLp2} on the sidelobe and mainlobe regions of the ACF expressed as
\begin{IEEEeqnarray}{rCl}
\text{GISR}~= \left(\dfrac{\int_{\Delta \tau}^{T}|R\left(\tau\right)|^p d\tau}{\int_{0}^{\Delta \tau}|R\left(\tau\right)|^p d\tau}\right)^{2/p}
\label{eq:GISR}
\end{IEEEeqnarray} 
where $p \geq 2$ is an integer.  When $p=2$, the GISR becomes the standard ISR metric.  As $p\rightarrow\infty$, the integrals in \eqref{eq:GISR} approach the infinity norm $||\cdot||_{\infty}$, also known as the max norm, and the GISR approaches the PSL metric \eqref{eq:PSL} \cite{bluntGISR}.  However, the max-norm tends to result in a discontinuous objective function which prevents the use of gradient-descent based waveform optimization methods.  Making $p$ large but finite results in a smooth objective function which allows for the use of gradient-descent based methods in the waveform optimization problem.  For this reason, the GISR metric will be the primary design metric this work uses to synthesize spectrally compact waveforms with desirably low ACF sidelobe levels. 


\section{CPM using the MTSFM Waveform Model}
\label{sec:MTSFM}
The MTSFM waveform's instantaneous phase is represented as a finite Fourier series expansion 
\begin{equation}
\varphi\left(t\right) = \frac{\alpha_0}{2} + \sum_{k=1}^K \alpha_k \sin\left(\frac{2\pi k t}{T}\right) + \beta_k \cos\left(\frac{2\pi k t}{T}\right)
\label{eq:MTSFM_1}
\end{equation}
where $\alpha_0$ is a constant phase term, $\alpha_k$ and $\beta_k$ are the Fourier coefficients, also known as modulation indices, and $K$ is the number of harmonics in the representation.  The MTSFM model has existed since at least the 1940's where it was used for the analysis of the spectra of arbitrary FM signals for use in analog communication systems \cite{Giacoletto}.  More recently, the author first utilized the MTSFM model to represent a class of Generalized Sinusoidal FM (GSFM) waveforms \cite{HagueDiss}.  Representing the modulation and phase functions of the GSFM waveform family allowed for deriving precise closed-form expressions for their spectra and AFs.  The MTSFM model can be used for waveform analysis in that it can represent any waveform whose modulation and phase functions meet the Dirichlet conditions \cite{Boyd}.  More recently, the author has utilized the MTSFM model for waveform synthesis where the modulation indices $\alpha_k$ and $\beta_k$ are modified to design waveforms with distinct AF/ACF properties \cite{Hague_AES, Hague_RadarConf17}.  This paper utilizes the MTSFM-CPM waveform model for both waveform analysis and synthesis.  The PC waveform's instantaneous phase is first approximated by a finite Fourier series.  This approximation is used as a form of CPM to smooth abrupt phase transitions in the PC waveform's instantaneous phase resulting in a more spectrally compact waveform.  The MTSFM-CPM waveform model is then used for waveform synthesis by modifying its modulation indices to refine the waveform's ACF sidelobe structure via the GISR metric defined in \eqref{eq:GISR}.  

The MTSFM-CPM waveform model in \eqref{eq:MTSFM_1} is used to approximate the instantaneous phase of a PC waveform in \eqref{eq:PC_Phase} by numerically computing the Fourier coefficients $\alpha_k$ and $\beta_k$.  As $K \to \infty$, the MTSFM-CPM approaches an exact representation of the PC waveform.  However, if $K$ is made finite, the Fourier series approximation to \eqref{eq:PC_Phase} is infinitely differentiable \cite{Boyd} and therefore smooth.  This smoothing of the PC waveform's instantaneous phase removes abrupt phase transitions and correspondingly improves the waveform's SC.  Reducing $K$ further smoothes the phase transitions and therefore further improves the waveform's SC.  To loosely derive a supremum on the number of harmonics $K$, consider the phase transition between two chips each of duration $t_b$.  The lowest frequency harmonic that can approximate this phase transition has a period $T_b = 2t_b$ and corresponding frequency $f_b = 1/2t_b$.  At a bare minimum then, the max frequency harmonic in the MTSFM, $K/T$, should be greater than or equal to $f_b$.  Noting that $T = N t_b$, setting $K/T \geq 1/2t_b$ and solving for $K$ results in the inequality
 \begin{IEEEeqnarray}{rCl}
K \geq \lceil N/2\rceil.
\label{eq:minK}
\end{IEEEeqnarray} 
The MTSFM-CPM approximation introduces perturbations to the PC waveform's instantaneous phase in the form of Gibbs phenomena.  These perturbations become substantial as $K$ approaches the limit \eqref{eq:minK} and degrades the waveform's AF/ACF sidelobe levels.  The MTSFM-CPM waveform's ACF properties must then be refined by modifying the modulation indices $\alpha_k$ and $\beta_k$. 

\section{Two Illustrative Design Examples}
\label{sec:designExamples}
This section demonstrates the MTSFM-CPM techique via two illustrative design examples.  The first PC waveform example utilizes a binary M-Sequence code and the second example uses a polyphase Barker code from \cite{Levanon}.

\subsection{M-Sequence Design Example}
\label{subsec:BPSK}

\begin{figure}[ht]
\centering
\includegraphics[width=0.5\textwidth]{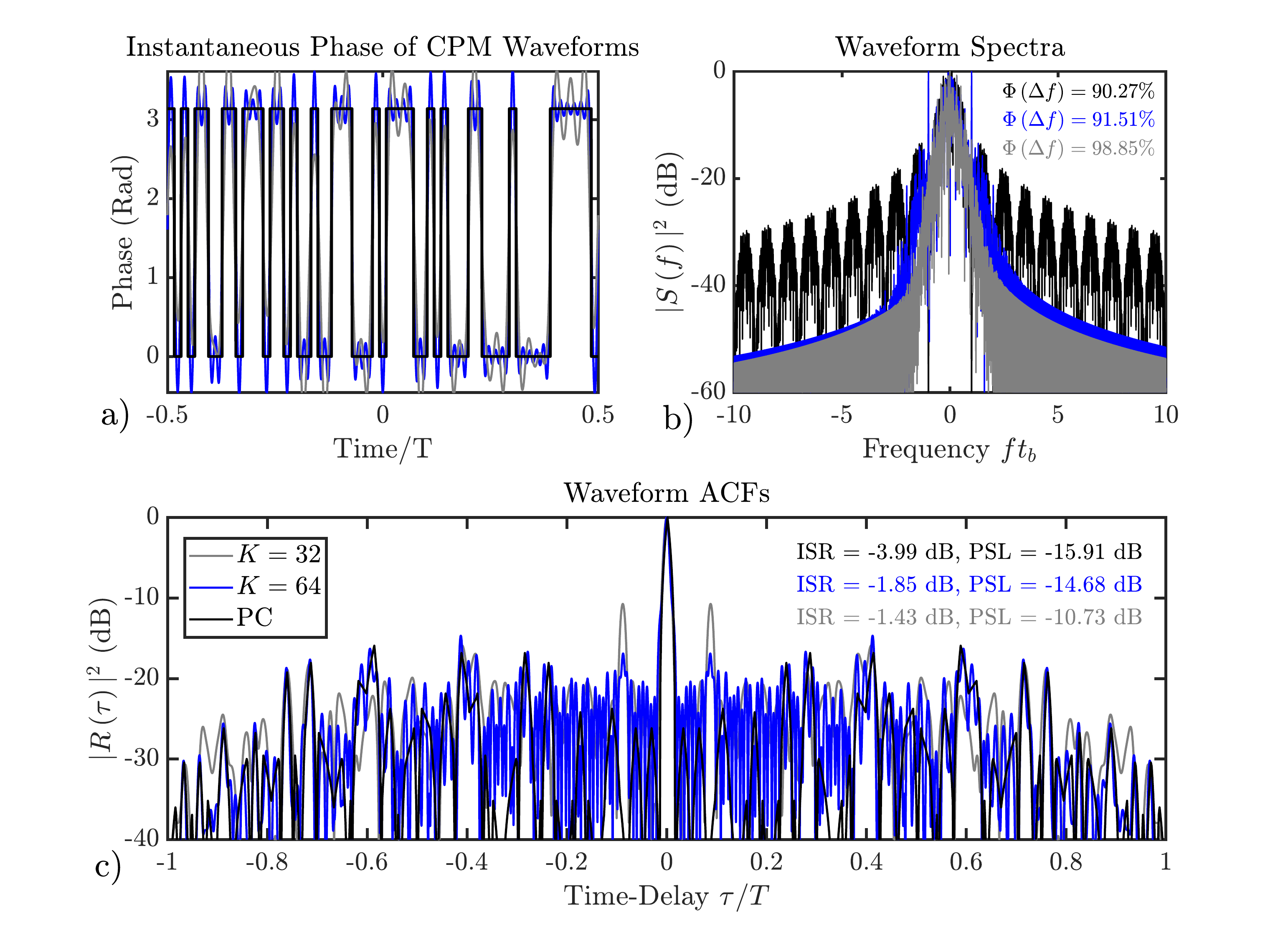}
\caption{Instantaneous phase (a), spectra (b), and ACFs (c) of the initial PC waveform with an M-Sequence phase code and MTSFM-CPM waveforms with $K=64$ and $32$ harmonics.  The MTSFM-CPM waveform smoothes the abrupt phase transitions of the PC waveform's instantaneous phase resulting in a more spectrally compact waveform at the cost of noticeably increased ACF sidelobes.}
\label{fig:MTSFM_1}
\end{figure}

The following design example demonstrates the MTSFM based CPM method for a PC waveform with $N=63$ chips utilizing an M-Sequence phase code \cite{Levanon}.  Figure \ref{fig:MTSFM_1} shows the instantaneous phases, spectra, and ACFs of the initial PC waveform and MTSFM-CPM waveforms composed of $K= 64$ and $32$ harmonics.  The original PC waveform possessed a SC \eqref{eq:THETA} of $90.27\%$ across the first spectral nulls (i.e, $\pm 1/t_b$) with ACF ISR and PSL values of -3.99 dB and -15.91 dB respectively.  The MTSFM-CPM waveforms possess spectra whose sidelobes fall off much more rapidly than the original PC waveform.  Their SC values across the same bandwidth as the PC waveform were $91.51\%$ and $98.85\%$ for $K= 64$ and $32$ harmonics respectively.  The higher SC values for these waveforms are a direct result of their increasingly smooth instantaneous phase functions.  However, the instantaneous phase of both MTSFM-CPM waveforms are perturbed by Gibbs phenomena, an artifact of the finite Fourier series representation of the PC waveform's instantaneous phase.  These perturbations in instantaneous phase degrade other properties of the waveform, specifically the ACF sidelobes.  The more spectrally compact MTSFM-CPM waveform with $K=32$ harmonics sees drastically increased ISR and PSL values of -1.43 dB and -10.73 dB respectively.  The MTSFM-CPM waveform is substantially more spectrally compact than the initial PC waveform at the expense of higher ACF sidelobes. 

\begin{figure}[ht]
\centering
\includegraphics[width=0.5\textwidth]{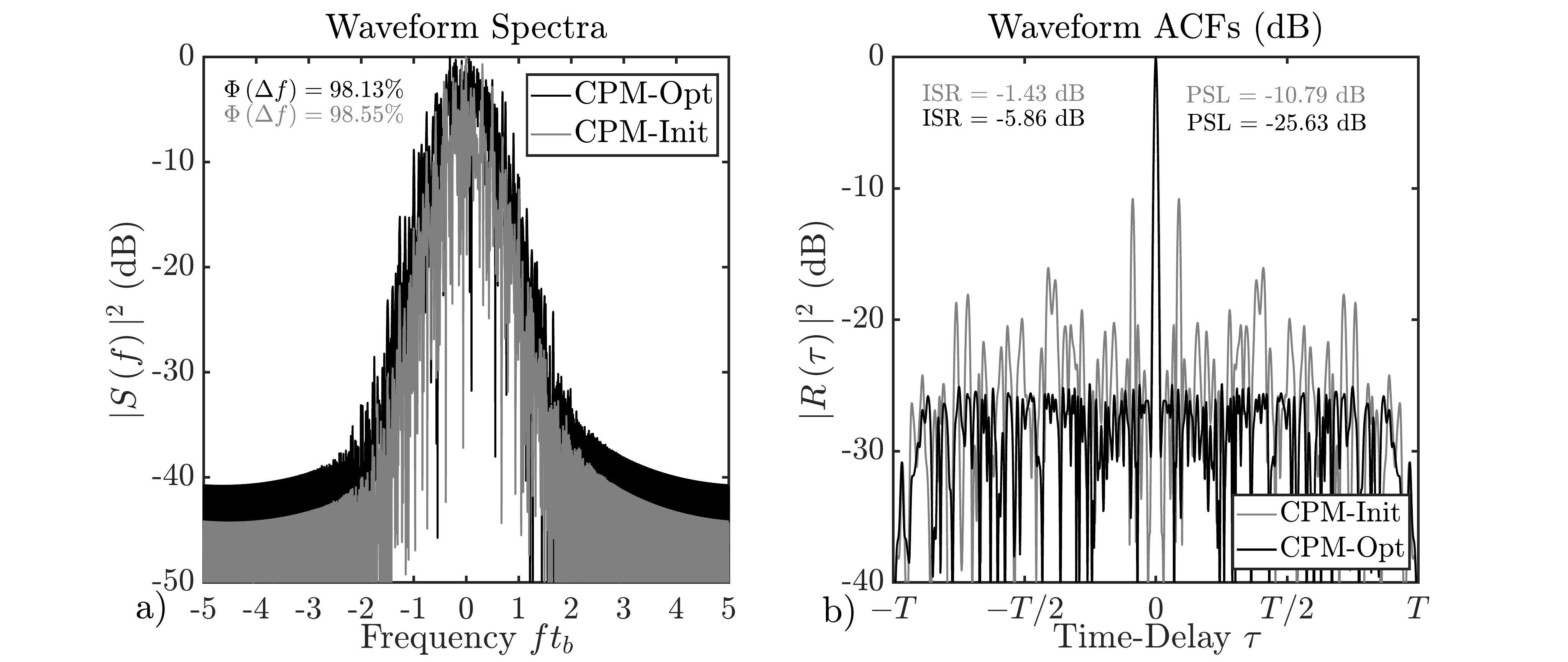}
\caption{Spectra (a) and ACFs (b) of the initial and optimized MTSFM-CPM waveforms.  The optimized MTSFM-CPM waveform possesses lower ACF sidelobes than both the initial MTSFM-CPM and PC waveform while also preserving the waveform's compact spectral shape.}
\label{fig:MTSFM_2}
\end{figure}

However, the MTSFM-CPM waveform's ACF properties can be refined.  This is achieved by modifying the modulation indices $\alpha_k$ and $\beta_k$ to optimize the waveform's GISR.  This re-optimization results in a MTSFM-CPM waveform with desireably low ACF sidelobes while retaining its compact spectral shape.  Formally, the optimization problem is defined as 
\begin{multline}
\underset{\alpha_{k}, \beta_k}{\text{min}}
\left(\dfrac{\int_{\Delta \tau}^T |R\left(\tau\right)|^p  d\tau}{\int_{0}^{\Delta \tau}|R\left(\tau\right)|^p d\tau} \right)^{2/p}  \text{s.t.~} \\ \beta_{rms}^2\left(\{\alpha_k,\beta_k\}\right) \in \left(1\pm\delta\right)\tilde{\beta}_{rms}^2
\label{eq:Problem1}
\end{multline}
where $\delta = 0.1$ is a unitless bound parameter and the GISR parameter $p=10$ favors a PSL optimization metric.  The $\tilde{\beta}_{rms}^2$ and $\beta_{rms}^2\left(\{\alpha_k, \beta_k\}\right)$ terms are the initialized and optimized MTSFM-CPM waveforms' RMS bandwidths respectively which are expressed in terms of $\alpha_k$ and $\beta_k$ as \cite{Hague_AES}
\begin{equation}
\beta_{rms}^2\left(\{\alpha_k, \beta_k\}\right) = \left(\frac{2\pi}{T}\right)^2 \sum_{k=1}^K k^2 \dfrac{\left(\alpha_k^2 + \beta_k^2\right)}{2}.
\label{eq:RMSBand}
\end{equation}
Recall from \eqref{eq:mainlobe} that the ACF mainlobe area is inversely proportional to RMS bandwidth \cite{Rihaczek} and thus the constraint on $\beta_{rms}^2$ will preserve the ACF mainlobe widith.  The \emph{fmincon} function in MATLAB's Optimization Toolbox \cite{Matlab} is used to minimize \eqref{eq:Problem1}.  This optimization function utilizes a Sequential Quadratic Programming (SQP) method in order to handle the nonlinear constraints in \eqref{eq:Problem1} and does not guarantee convergence to a global minimum.  Figure \ref{fig:MTSFM_2} shows the spectra and ACFs of the initial PC waveform as well as the initial and optimized MTSFM-CPM waveforms.  The optimized MTSFM-CPM waveform retains a compact spectral shape with a SC of $98.13\%$ while also possessing substantially lower ACF sidelobes than the initial MTSFM-CPM waveform.  The ISR and PSL values for this optimized MTSFM-CPM waveform are $-5.86$ dB and $-25.63$ dB respectively.  The optimized MTSFM-CPM waveform is not only more spectrally compact than the original PC waveform, but also possesses substantially lower ACF sidelobes than it as well.  

\subsection{Polyphase Barker Code Design Example}
\label{subsec:polyphase}
\begin{figure}[ht]
\centering
\includegraphics[width=0.5\textwidth]{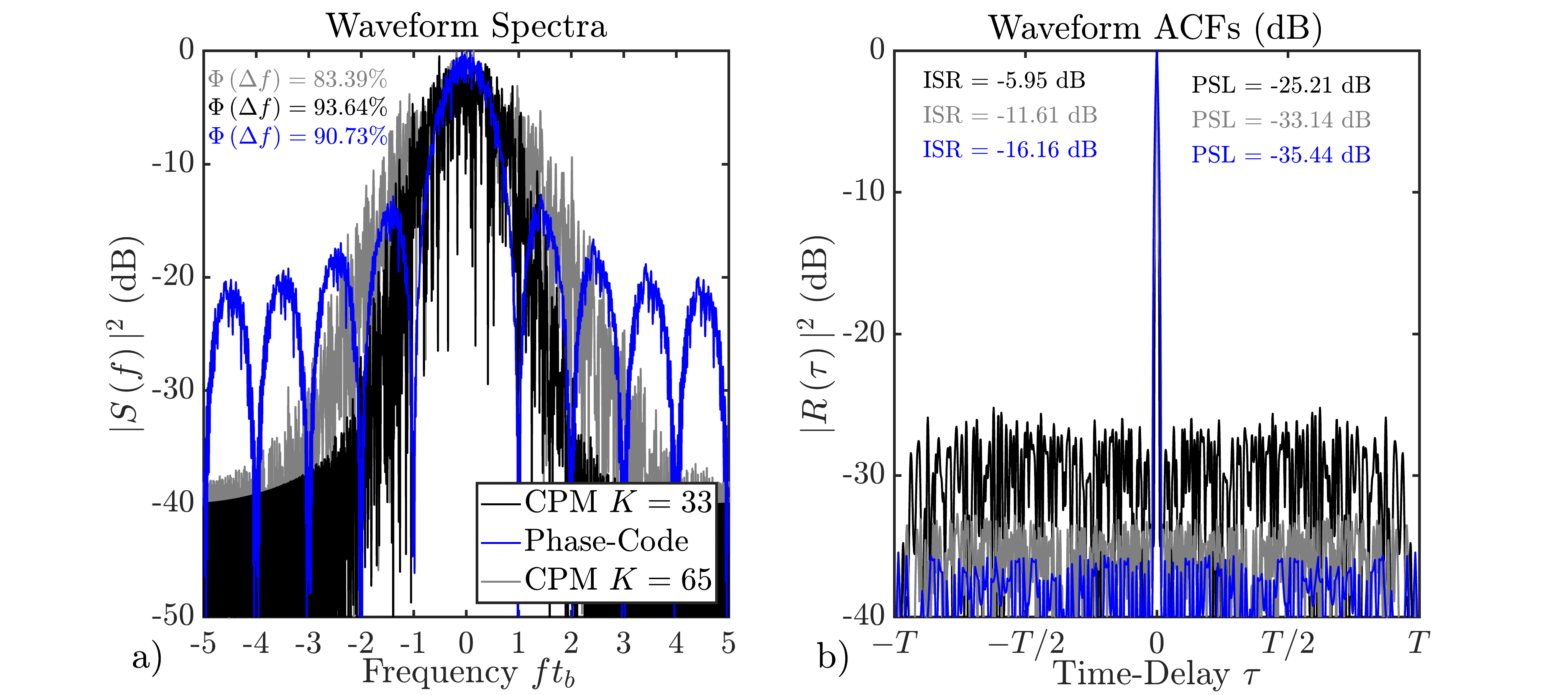}
\caption{Spectra (a) and ACFs (b) of original PC waveform and optimized MTSFM-CPM waveforms with $K = 33$ and $65$ harmonics.  The MTSFM-CPM waveform with $K=33$ harmonics possesses better SC but higher sidelobes than the $K=65$ harmonic MTSFM-CPM waveform.  Neither MTSFM-CPM waveform achieves the same low sidelobe levels as the original PC waveform. }
\label{fig:MTSFM_3}
\end{figure}

The results from the previous design example showed that re-optimizing the MTSFM-CPM waveform's modulation indices produced a spectrally compact waveform with ACF sidelobes that were lower than the initial PC waveform from which the MTSFM-CPM was derived.  However, the following design example shows that such a result is not always possible.   Figure \ref{fig:MTSFM_3} shows the spectra and ACFs of a PC waveform with $N=65$ chips employing a polyphase Barker code and two optimized MTSFM-CPM waveforms derived from it with $K=65$ and $K=33$ harmonics representing their instantaneous phase.  While the MTSFM-CPM waveform with $K=33$ harmonics was clearly the most spectrally compact, its ACF sidelobes were more than 10 dB higher than the initial PC waveform's ACF sidelobes.  The MTSFM-CPM waveform with $K=65$ harmonics achieved much lower ACF sidelobes and were closer to that of the initial PC waveform's ACF sidelobes.  This waveform also possessed a spectrum whose sidelobes fell off more rapidly with increasing frequency than the inital PC waveform.  However, it did not concentrate as much of its spectral energy as densely as the initial PC waveform with a SC of  $83.39\%$ computed over the PC waveform's null-to-null bandwidth.  Increasing $K$ produced a MTSFM-CPM waveform with lower ACF sidelobes at the expense of reduced SC.  This example shows that the choice of $K$ in the MTSFM-CPM introduces a tradeoff between SC and low ACF sidelobes.

\section{Conclusion}
\label{sec:Conclusion}
This paper describes a CPM technique for improving on a PC waveform's SC properties utilizing the MTSFM-CPM waveform model.  The PC waveform's instantaneous phase is represented as a finite Fourier series which smooths its abrupt phase transitions resulting in a more spectrally compact waveform.  This smoothing of the PC waveform's instantaneous phase introduces perturbations to it which degrades the resulting MTSFM-CPM waveform's ACF sidelobe structure.  Adjusting the MTSFM-CPM waveform's modulation indices refines its ACF sidelobes while also largely preserving its compact spectral shape.  Adjusting the number of harmonics $K$ in the MTSFM-CPM model facilitates trading off between SC and preserving the original PC waveform's low ACF sidelobe levels.  A lower $K$ produces MTSFM-CPM waveforms with higher SC but increased ACF sidelobe levels.  A higher K produces MTSFM-CPM waveforms with lower ACF sidelobes at the expense of degraded SC.  Future efforts will focus on a more through analysis of the tradeoff between SC and ACF sidelobe levels through the choice of the number of harmonics $K$ for a wide variety of phase code lengths $N$.  Also of interest is extending this method to designing families of spectrally compact waveforms with desireable Auto and Cross-Correlation Function properties.  

\section*{Acknowledgment}
The author would like to thank Professor John R. Buck at the University of Massachusetts Dartmouth for the many insightful converstations during the author's dissertation studies that inspired this work. 


\end{document}